\documentclass[11pt]{article}

\usepackage[margin=1in]{geometry}
\usepackage{amsmath,amssymb,amsthm,mathtools}
\usepackage{booktabs}
\usepackage{enumitem}
\usepackage{authblk}
\usepackage{xcolor}
\usepackage[numbers,sort&compress]{natbib}
\usepackage[colorlinks=true,linkcolor=blue!50!black,citecolor=blue!50!black,urlcolor=blue!50!black]{hyperref}
\newcommand{\resultlabel}[2]{\par\medskip\noindent\textbf{#1 (#2).}\ }

\title{\bfseries Thermodynamic Realizability of Hidden Markov Processes\\[4pt]
\large Attainment, Observable Certificates, an Architecture-Independent Floor,\\ and the Price of Architecture}

\author{Murad Aznagulov}
\affil{Lomonosov Moscow State University \\ \texttt{amurad12@icloud.com}}

\date{September 2026; revised November 2026 (v2)}

\begin{document}
\maketitle

\begin{abstract}
\noindent
What is the least dissipative finite Markov machine that can reproduce a given stochastic process \emph{exactly}? A basic compactness worry appears immediately: a sequence of cheaper and cheaper models might lower its cost only by sending microscopic rates to infinity, never settling into an actual finite-rate machine. We show this escape is impossible at any fixed hidden dimension. States whose stationary occupation vanishes can be traced out, and clusters of transitions running arbitrarily fast can be contracted into a single effective mode, in neither case raising entropy production or losing the limiting observed path law. The exact-law cost $V_n(P)$ is therefore attained and lower semicontinuous at every fixed state count, with no cap on rates or on mean activity. Charging a positive price $\kappa$ per used hidden state turns this into a genuine design problem: an optimum now exists over \emph{all} finite dimensions at once, with an exact representation as a supremum of globally calibrated lower certificates built from finitely many observable path expectations. The resulting lower envelope $G_\kappa(P)=\min_n[V_n(P)+\kappa n]$ traces out an architecture phase diagram, whose slope is the state count the price selects; that count diverging as $\kappa\downarrow0$ turns out to be exactly equivalent to no finite machine ever attaining the all-dimension infimum $V_\infty(P)$. We then bound the phase diagram from below, uniformly in architecture. Every Markov jump process has a reversible companion, with the same states, stationary law and edge set, whose path law is within relative entropy $T\sigma/4$ of the original on every window of length $T$; the constant is sharp. Combined with a Loewner-order estimate for the sojourn cycle operators of reversible chains, this gives $\sigma\ge4r_C\,D_{\mathrm{CM}}(f_C)$ for every finite realization of any size. Here $r_C$ is the visit rate of an observed set $C$, $f_C$ its sojourn-time density, and $D_{\mathrm{CM}}$ the relative entropy from $f_C$ to the cone of completely monotone densities, a convex program with a Donsker--Varadhan dual certificate. For the time-symmetric two-block renewal family $P_q$ we derive the complete phase-type cancellation equations, exhibit an explicit four-phase system that defeats a restricted parameterization, and prove positive dissipation at every finite dimension. The new floor gives $V_\infty(P_q)>0$ for all $0<q<1$, with $V_\infty(P_{1/20})\ge0.1025$. This retracts the conjecture $V_\infty(P_{1/20})=0$ made in v1 and excludes the complexity-assisted approach to reversibility. Exact rational realizations with four to six phases show that the three-state machine is not optimal: $V_7(P_{1/20})\le2.73611<(1-q)\log(1/q)$. The question left open is whether the optimal architecture saturates at a finite size or diverges toward a positive floor.
\end{abstract}

\section{The question: can a physical optimum disappear?}

A hidden Markov model can explain what is observed without being the cheapest physical system that could have produced it. Two different knobs can lower the apparent cost: adding hidden states, and letting some transition rates run away to infinity. These are genuinely different mechanisms --- one changes the architecture, the other changes the clock --- and a theory of physical cost has to keep them apart.

This paper builds that theory. At a fixed number of hidden states, we show that letting rates diverge cannot make the cheapest exact realization disappear: a minimizing sequence of ever-faster machines still converges to an actual, finite machine of no greater cost. Charging a positive price per hidden state then extends this from a fixed-dimension statement to an unrestricted one: the same attainment holds with no ceiling on dimension at all, and the resulting optimum has an exact representation as a supremum of continuous, observable lower certificates.

This is a statement about physically \emph{realizing} a stationary process --- the machine that produces the data --- not about the thermodynamic cost of a computer that merely \emph{predicts} it. These are distinct optimization problems, and no identification of memory cost with heat is made here.

This problem sits next to a growing literature on bounding entropy production from partial observations and hidden Markov structure \citep{Ehrich2021,SkinnerDunkel2021,NitzanGhosalBisker2023,MaierSeifertVanderMeer2025,Seifert2026Review}. Coarse-graining and fast-state elimination are older, classical themes \citep{PuglisiEtAl2010,Jia2016}, and in a different master-equation setting hidden states are known to change the resources needed to implement a prescribed stochastic map \citep{OwenKolchinskyWolpert2019}. Closest in spirit to the mechanism used here, Dechant showed that a prescribed time evolution can be implemented at arbitrarily small entropy production if one is willing to pay with diverging activity \citep{Dechant2022} --- essentially the opposite trade from the one our compactness theorem rules out. What we ask is narrower and, we think, cleaner: fix the \emph{entire stationary observed path law} exactly, leave microscopic rates completely unrestricted, and optimize only over finite hidden continuous-time Markov realizations. Under that exact-law constraint, it turns out that diverging activity alone cannot destroy a fixed-dimension optimum.

The main statements are:

\begin{center}
\begin{tabular}{@{}p{0.52\textwidth}p{0.4\textwidth}@{}}
\toprule
\textbf{Question} & \textbf{Answer proved here} \\
\midrule
Can a fixed-dimension minimizing sequence escape through infinite rates? &
Its limit can be represented by a finite CTMC of no larger dimension and no greater cost. \\[4pt]
Is the optimum stable under perturbations of the observed path law? &
Yes at fixed dimension, in the specified path topology. \\[4pt]
Does a positive state-count price give a finite optimal architecture? &
Yes, without a rate or activity cap. \\[4pt]
Can observable certificates recover that priced optimum exactly? &
Yes, as a supremum of globally calibrated continuous certificates. \\[4pt]
Can hidden architecture drive the dissipation of a time-symmetric law to zero? &
Not if some observed sojourn law is not completely monotone: a floor uniform in size (Theorem D). \\[4pt]
Is $V_\infty(P_q)$ positive? &
Yes, for every $0<q<1$; $V_\infty(P_{1/20})\ge0.1025$ (Corollary D.1). \\[4pt]
Is the three-state machine for $P_{1/20}$ optimal? &
No; exact six-phase realization with $\sigma<2.7362$ (Proposition 11). \\[4pt]
Is every finite realization of the target renewal law improvable? &
This remains open; it is not implied by the preceding results. \\
\bottomrule
\end{tabular}
\end{center}

The last row is the one to watch. Everything above it is proved; the last row is not. We establish a general variational foundation, bound it from below uniformly in the architecture, and isolate exactly what remains open --- the all-dimension nonattainment question --- rather than quietly obtaining an unproved numerical global optimum by renaming it.

\paragraph{Changes in v2.} Version~1 conjectured $V_\infty(P_{1/20})=0$ (former \S\ref{sec:scen}). That conjecture is false: Theorem~D and Corollary~D.1 (new \S\ref{sec:floor}) prove an architecture-independent positive floor. The other new material is Lemma~D.0 (reversible companion), Proposition~10 (the floor as an explicit observable certificate for the phase diagram), Proposition~11 (exact multi-phase witnesses), Table~1, and the extended ancillary code. The numbering of all v1 results is unchanged. A companion work~\cite{AznagulovTwoPrices} treats the memory cost of detailed balance and proves, by a total-variation route, the near-equilibrium scaling $V_\infty(P_q)=\Theta((1-q)^2)$; the relative-entropy floor here is sharper far from equilibrium (Table~1).

\section{Definitions and how to read the proofs}

We fix notation precisely in this section; a reader willing to take entropy production and the cost functional $V_n$ on faith can skim it and return only as needed.

For a finite irreducible continuous-time Markov chain with generator $Q$ and stationary row vector $\pi$, write
\[
F_{ij}=\pi_i Q_{ij}, \qquad
\sigma(Q)=\sum_{i<j} J(F_{ij},F_{ji}), \qquad
K(Q)=\sum_i \pi_i(-Q_{ii}).
\]
For positive arguments, the edge cost is $J(x,y)=(x-y)\log(x/y)$. Set $J(0,0)=0$ and $J(x,0)=J(0,x)=+\infty$ when $x>0$. The same entropy rate can be written
\[
\sigma=\sum_{i\neq j} F_{ij}\log(F_{ij}/F_{ji}).
\]
We use ordinary time reversal of even variables and set Boltzmann's constant to one. Mathematically, $\sigma$ is the stationary path-space irreversibility rate of the finite jump process; under the standard local-detailed-balance interpretation of Markov-jump stochastic thermodynamics it is the steady entropy-production rate. The quantity $K$ counts stationary microscopic jumps per unit time. It is distinct from entropy production and from the observed jump count.

A realization includes a deterministic observation map $h$ into a fixed finite alphabet. It \emph{realizes} $P$ when the entire stationary observed path law is $P$. Define
\[
V_n(P)=\inf\{\sigma(Q):\ |Q|\le n,\ P_{Q,h}=P\}, \qquad V_\infty(P)=\inf_n V_n(P).
\]
Empty infima are infinite. The use of ``at most $n$'' avoids counting unused states. For unrestricted rates, balanced splitting also permits padding to an exact dimension without changing law or entropy production.

In the earlier renewal calculations, $N$ counted the phases of block $B$, with one additional reset state $A$. Thus that $N$ corresponds to $n=N+1$ in the total-state convention here. The class with a single reset state is a subclass of arbitrary hidden CTMC realizations. We do \emph{not} assume that its optimum equals the unrestricted hidden-model optimum. Corollary 5.1 proves fixed-$N$ attainment within the reset class only.

Theorems A--D are the main results. The numbered technical results collected afterwards supply the compactification proof and the renewal-law applications. All state counts in the main theorems count total microscopic states.

\section{Closure when the observed law also changes}

The earlier compactification used the same observed law along a sequence. Observable certificate representation needs a stronger statement: the laws themselves may approach the target, not merely a fixed target approached by a sequence of machines.

Let $\mathcal{E}$ denote stationary ergodic laws on a fixed finite observation alphabet with c\`{a}dl\`{a}g paths. Equip this set with \emph{local weak Skorokhod $J_1$ convergence}: on every finite time window whose endpoints are almost surely continuity times of the limiting path, the path distributions converge weakly. Stationarity makes every deterministic endpoint a continuity time almost surely. This topology retains information about short visible excursions, even when their duration tends to zero.

\resultlabel{Theorem A}{observable-law closure at bounded dimension}
Suppose finite irreducible stationary CTMCs $Q_k$ have at most $n$ states, their observed laws $P_k$ converge in this topology to $P\in\mathcal{E}$, and
\[
L=\liminf_k \sigma(Q_k) < \infty.
\]
Then $P$ has a finite irreducible exact realization with at most $n$ states and entropy production at most $L$. Consequently $V_n$ is lower semicontinuous on $\mathcal{E}$.

\begin{proof}
Repeat the compactification of Theorem~5, but note one additional point: when the observed laws vary with $k$, tracing out states of vanishing stationary mass must not erase a visible excursion that survives in the limit. The trace clock estimate (4.3) is unchanged, so it is enough to prove a uniform short-window bound for visible jumps.

Fix a deterministic time $t$ and a finite window containing $[t-1,t+1]$ in its interior. By Skorokhod representation, after passing to a coupled version the observed paths $Y^{(k)}$ converge almost surely to $Y$ in $J_1$ on this window. A c\`{a}dl\`{a}g path with values in a finite discrete alphabet is locally constant at every continuity time. Stationarity implies that a deterministic $t$ is a continuity time of $Y$ almost surely. For any $\eta>0$, on the event that $Y$ is constant on $[t-2\eta,t+2\eta]$, the defining time changes in $J_1$ convergence are eventually within $\eta/2$ of the identity and the spatial discrepancy is eventually smaller than the minimum nonzero distance between observation symbols. Hence $Y^{(k)}$ is eventually constant on $[t,t+\eta]$. Therefore
\[
\limsup_{k\to\infty}\Pr_{P_k}\{\text{a jump in }[t,t+\eta]\}
\le
\Pr_P\{\text{a jump in }[t-2\eta,t+2\eta]\}.
\]
The right-hand side tends to zero as $\eta\downarrow0$, because a finite-alphabet c\`{a}dl\`{a}g path has only finitely many jumps on compact intervals and has no jump at deterministic $t$ almost surely. Thus
\begin{equation*}
\lim_{\eta\downarrow0}\ \limsup_k \Pr\nolimits_{P_k}\{\text{a jump in } [t,t+\eta]\} = 0. \tag{A.1}
\end{equation*}
One-sided neighborhoods suffice at an endpoint.

Now trace away states whose stationary mass tends to zero. Combining (A.1) with (4.3), for any finite collection of observation times the time change induced by tracing vanishes in probability and the probability of crossing a visible jump during that time change vanishes uniformly in $k$. Hence the observed finite-dimensional distributions of the traced chains still converge to those of $P$. Their positive limiting stationary masses are bounded away from zero. The conductance partition, the semigroup collapse, the monochromatic-block argument, and the entropy log-sum estimate from Theorem~5 then apply verbatim and produce a finite exact realization of $P$ with cost at most $L$.

For lower semicontinuity, let $P_k\to P$ and choose realizations with cost at most $V_n(P_k)+1/k$ along a subsequence realizing the $\liminf$. If that $\liminf$ is finite, the first part gives $V_n(P)\le\liminf_kV_n(P_k)$; if it is infinite the inequality is immediate.
\end{proof}

\paragraph{Why this topology is specified.} Convergence of a few sampled correlations, moments, or even arbitrarily short individual holding times is not the hypothesis. A path that makes an excursion to another symbol and immediately returns does not converge in $J_1$ to the path with that excursion erased. This prevents the compactification from silently discarding visible events. We do not claim this theorem under the weaker rational-cylinder topology alone.

The zero-cost characterization is now robust.

\resultlabel{Corollary A.1}{equilibrium realizability at finite dimension}
\begin{equation*}
V_n(P)=0 \iff P \text{ has a reversible realization with at most } n \text{ states.}
\end{equation*}
If $V_n(P)>a$, some neighborhood of $P$ in the specified topology satisfies $V_n(R)>a$ throughout.

\begin{proof}
Attainment and the equality condition in $J$ prove the equivalence. Lower semicontinuity makes the strict superlevel set open.
\end{proof}

Thus the positive fixed-dimension gap for $P_{1/20}$ also survives sufficiently small path-law perturbations. This supplies existence of a robust neighborhood, not a numerical radius or a finite-sample confidence interval.

\section{The master theorem with a price for architecture}

Let $\kappa>0$ be a price per used microscopic state, in the same objective units as entropy production. This is a modeling parameter; it is not claimed to be a universal physical energy-per-state constant. Define
\begin{equation*}
G_\kappa(P) = \inf_{Q:\ P_Q=P} \{\sigma(Q)+\kappa|Q|\}. \tag{B.1}
\end{equation*}
The class is all finite irreducible stationary CTMCs with deterministic observation. Assume the target has at least one finite-cost realization.

Choose a countable family $\phi_j$ of bounded continuous finite-window path functionals, normalized to $[0,1]$, that determines the local weak topology on stationary laws. Such a family exists from separability of the Skorokhod path spaces, using a countable convergence-determining family on each integer window. Set
\begin{equation*}
z_j(P)=\mathbb{E}_P\phi_j, \qquad
d(P,R)=\sum_{j\ge1} 2^{-j}|z_j(P)-z_j(R)|, \qquad
d_m=\sum_{j=1}^m 2^{-j}|z_j(P)-z_j(R)|. \tag{B.2}
\end{equation*}
The observables are path functionals on finite windows, not necessarily raw jump counts or probabilities at isolated time points.

For $L>0$ define a globally calibrated certificate
\begin{equation*}
B_{\kappa,L,m}(P) = \inf_Q \{\sigma(Q)+\kappa|Q|+L\,d_m(P,P_Q)\}. \tag{B.3}
\end{equation*}
For fixed $L,m$, the triangle inequality implies
$|B_{\kappa,L,m}(P)-B_{\kappa,L,m}(R)|\le Ld_m(P,R)$, so each certificate is continuous and depends only on the first $m$ observable expectations.

\resultlabel{Theorem B}{attainment, stability, and exact observable representation}
For every $\kappa>0$:
\begin{equation*}
G_\kappa(P) = \min_{n\ge1}\{V_n(P)+\kappa n\} = \sup_{L>0,\ m\ge1} B_{\kappa,L,m}(P). \tag{B.4}
\end{equation*}
The physical minimum is attained by a finite irreducible CTMC, $G_\kappa$ is lower semicontinuous on $\mathcal{E}$, and each certificate is continuous and depends on finitely many observable expectations. No rate or mean-activity constraint is imposed.

\begin{proof}[Proof of attainment]
A finite competitor gives an upper bound $C$ on the objective. A minimizing sequence then has at most $(C+1)/\kappa$ states. Theorem~5 gives an exact limiting realization of no greater entropy and no greater state count, proving attainment.

For the first equality, a realization using exactly $m$ states has cost at least $V_m+\kappa m$, so $G_\kappa\ge\inf_m(V_m+\kappa m)$. Conversely, if an optimizer attaining $V_n$ actually uses $m\le n$ states, then $V_m\le V_n$ while monotonicity gives $V_n\le V_m$, hence $V_m=V_n$; its priced cost is therefore $V_m+\kappa m$. Taking the best such used count gives the reverse inequality. The displayed infimum is a minimum by the same state-count bound.
\end{proof}

\begin{proof}[Proof of stability]
If $P_k\to P$ and the $\liminf$ of $G_\kappa(P_k)$ is finite, choose nearly optimal realizations. Their counts are bounded. Fix the count along a subsequence and apply Theorem A. Entropy and state count do not increase, proving lower semicontinuity.
\end{proof}

\begin{proof}[Proof of the observable representation]
Put
\[
B_{\kappa,L}(P) = \inf_Q \{\sigma(Q)+\kappa|Q|+L\,d(P,P_Q)\}.
\]
It is $L$-Lipschitz and no larger than $G_\kappa$. Its supremum over $L$ is therefore a lower-semicontinuous minorant.

Conversely, if $b=\sup_L B_{\kappa,L}(P)<\infty$, choose $Q_L$ within $1/L$ of the infimum for integer $L$. Nonnegativity gives
\[
d(P,P_{Q_L}) \le (b+1)/L, \qquad \sigma(Q_L)+\kappa|Q_L| \le b+1/L.
\]
State counts are bounded. Theorem A applied to this sequence implies $G_\kappa(P)\le b$. Infinite $b$ is immediate.

Finally
\[
B_{\kappa,L,m} \le B_{\kappa,L} \le B_{\kappa,L,m} + L\,2^{-m},
\]
which proves (B.4).
\end{proof}

This is an exact nonlinear observable dual representation. It is not a claim of a finite-dimensional linear-program dual or an efficient calibration algorithm. A locally fitted generator produces an upper approximation to the infimum in (B.3) and is not a valid lower certificate. Global calibration remains a substantive task.

There is a corresponding fixed-dimension formula:
\begin{equation*}
V_n(P) = \sup_{L>0,\,m\ge1}\ \inf_{|Q|\le n} \{\sigma(Q)+L\,d_m(P,P_Q)\}. \tag{B.5}
\end{equation*}
The proof uses Theorem A with the dimension ceiling already imposed.

If simultaneous estimation guarantees $|\hat z_j-z_j(P)|\le\epsilon_j$ for $j\le m$, then
\begin{equation*}
G_\kappa(P) \ge B_{\kappa,L,m}(\hat z) - L\sum_{j=1}^m 2^{-j}\epsilon_j \tag{B.6}
\end{equation*}
on that event. The theorem supplies the deterministic transfer of errors; it does not supply independence or mixing assumptions for a particular data record.

\section{The architecture phase diagram and the unpriced limit}

For a fixed finite-realizable $P$, regard $G_\kappa(P)$ as a function of the state price.

\resultlabel{Theorem C}{architecture selected by its price}
\begin{enumerate}[label=(\arabic*)]
\item $G_\kappa$ is nondecreasing, concave, and piecewise affine locally on $\kappa>0$. On every compact positive price interval, only finitely many dimensions can be optimal.
\item If $\kappa_1<\kappa_2$, every optimal used count $n_1$ at $\kappa_1$ and $n_2$ at $\kappa_2$ satisfies $n_1\ge n_2$.
\item At a differentiability point, $G_\kappa'=n_\kappa$, the common optimal count. The right derivative is the smallest optimal count and the left derivative the largest.
\item Writing $V_\infty=\inf_n V_n$,
\begin{equation*}
V_\infty(P) = \lim_{\kappa\downarrow0} G_\kappa(P) = \lim_{\kappa\downarrow0}\ \sup_{L,m} B_{\kappa,L,m}(P). \tag{C.1}
\end{equation*}
\item Every optimal count $n_\kappa$ tends to infinity as $\kappa\downarrow0$ if and only if no finite realization attains $V_\infty$. If a finite all-dimension optimizer exists and its smallest used count is $n^*$, then all sufficiently small positive prices select exactly $n^*$ and entropy $V_\infty$.
\end{enumerate}

\begin{proof}
Equation (B.4) is a lower envelope of affine functions with positive integer slopes. A fixed competitor and a positive lower bound on $\kappa$ give a uniform bound on the active counts, proving the first and third assertions.

Compare the optimality inequalities for two selected realizations:
\[
\sigma_1+\kappa_1 n_1 \le \sigma_2+\kappa_1 n_2, \qquad \sigma_2+\kappa_2 n_2 \le \sigma_1+\kappa_2 n_1.
\]
Adding proves $(\kappa_2-\kappa_1)(n_2-n_1)\le0$.

The lower bound $G_\kappa\ge V_\infty$ is immediate. For any finite competitor of entropy below $V_\infty+\varepsilon$, its finite dimension gives $\limsup_{\kappa\downarrow0}G_\kappa \le V_\infty+\varepsilon$; let $\varepsilon\downarrow0$.

If optimal counts remain bounded along a sequence of prices tending to zero, their entropies tend to $V_\infty$, and fixed-dimension attainment supplies a finite all-dimension optimizer. Conversely, a smallest all-dimension optimizer of count $n^*$ gives $\sigma_\kappa+\kappa n_\kappa \le V_\infty+\kappa n^*$. Since $\sigma_\kappa\ge V_\infty$, we have $n_\kappa\le n^*$. Each feasible smaller dimension has a strictly positive gap above $V_\infty$, by minimality and attainment. There are finitely many such dimensions, so sufficiently small prices exclude all of them. The optimal count is then $n^*$ and its entropy must equal $V_\infty$.
\end{proof}

Equation (C.1) is a full exact-cost representation with an ordered architecture-price limit, but that ordering is not a license to swap the limit with the supremum over observable certificates: at zero price we prove no closure theorem for the union over all finite dimensions, so that exchange is left unproved here.

This reformulates the all-$N$ question as a definite physical phase question: as the price of architecture vanishes, does the chosen architecture eventually stop growing? The answer for $P_{1/20}$ remains unknown. Neither fixed-dimension attainment, positive dissipation, nor the regularized certificate representation decides it. What can now be decided is the \emph{height} of the zero-price limit: Theorem~D bounds $V_\infty=\lim_{\kappa\downarrow0}G_\kappa$ from below by an explicit, architecture-independent observable functional (Proposition~10).

\section{What the architecture price reveals --- and what remains open}\label{sec:scen}

The regularized problem separates three physically distinct zero-price scenarios. This distinction is useful because only the first is compatible with a finite thermodynamically final architecture.

\begin{enumerate}[label=(\roman*)]
\item \textbf{Finite saturation.} Some finite $n^*$ attains $V_\infty$. Then Theorem C implies that all sufficiently small prices select exactly $n^*$ states. Hidden architecture eventually stops buying thermodynamic improvement.
\item \textbf{Divergent architecture with a positive floor.} No finite realization attains $V_\infty>0$. Then $n_\kappa\to\infty$ while $G_\kappa\to V_\infty>0$. Increasing microscopic complexity continues to help forever, but an irreducible positive dissipation floor remains.
\item \textbf{Complexity-assisted approach to reversibility.} $V_\infty=0$ while every finite $V_n>0$. Then necessarily $n_\kappa\to\infty$ and $G_\kappa\to0$. In this strongest regime, no finite machine is reversible, yet arbitrarily small dissipation is approached by increasing hidden architecture.
\end{enumerate}

Theorem 9 rules out a finite zero-cost realization of the renewal family $P_q$, $0<q<1$, but by itself leaves all three scenarios open. Version~1 of this paper conjectured the strongest one, $V_\infty(P_{1/20})=0$. \emph{That conjecture is false.} Theorem~D and Corollary~D.1 below give $V_\infty(P_q)>0$ for every $0<q<1$, and $V_\infty(P_{1/20})\ge0.1025$. Scenario~(iii) is therefore excluded for the whole family. Hidden architectural complexity is not a substitute for dissipation, even asymptotically. What remains is the dichotomy between (i) and (ii). By Theorem~8 it is equivalent to the question whether every finite realization can be strictly improved. Proposition~11 shows that the simplest candidate, the three-state machine, can be.

More generally, Theorem~D gives a necessary condition for scenario~(iii) for \emph{any} target law: every observed sojourn-time law must be completely monotone. Laws that pass this test can still fail to be finitely reversible because of hidden memory carried across symbol changes \cite{AznagulovTwoPrices}; for such laws the zero-price limit is a genuinely open problem.

The price variable also turns asymptotic dimension gaps into observable scaling laws for the optimization problem. For example, if future work establishes
\[
V_n-V_\infty \sim A n^{-\alpha}\qquad(A,\alpha>0),
\]
then minimizing $A n^{-\alpha}+\kappa n$ predicts
\[
n_\kappa\sim\left(\frac{\alpha A}{\kappa}\right)^{1/(\alpha+1)},
\qquad
G_\kappa-V_\infty\asymp \kappa^{\alpha/(\alpha+1)}.
\]
This is a conditional consequence, not a measured exponent. It suggests a concrete research program: certify pieces of the $\kappa$--architecture phase diagram, infer the asymptotic regime, and only then attack the zero-price limit analytically.

Two extensions would make the framework more directly physical. First, a device model could replace the abstract state penalty $\kappa|Q|$ by a justified maintenance or fabrication power for microscopic degrees of freedom. Second, one could replace deterministic observation by a noisy observation channel or move from finite CTMCs to semi-Markov and diffusion realizations. The compactness mechanism identifies exactly what would have to be re-proved in those settings: control of disappearing mass, fast internal modes, preservation of the observed law, and lower semicontinuity of dissipation.

\section{Hidden architecture cannot buy reversibility: an architecture-independent floor}\label{sec:floor}

The results so far are uniform in rates but not in dimension. This section supplies the missing uniformity in dimension. The difficulty is real: for a time-symmetric observed law every lower bound built from the asymmetry between the observed path law and its reversal vanishes, while the hidden machine may be arbitrarily large.

\paragraph{Sojourn data.} Let $P$ be stationary and ergodic and $C$ a nonempty proper subset of the observation alphabet. A $C$-sojourn is a maximal time interval on which the observed path stays in $C$. Let $r_C\in(0,\infty)$ be the mean number of $C$-sojourns started per unit time and $f_C$ the Palm density of their durations. For $P_q$ with $C=\{B\}$, $r_C=r=s/3$ and $f_C=f_q$. A density on $(0,\infty)$ is \emph{completely monotone} if it equals $\int\lambda e^{-\lambda u}\nu(d\lambda)$ for a probability measure $\nu$ on $(0,\infty)$ \cite{Widder1941}. By (9.2), every complete dwell density of a finite reversible realization has this form, as is classical \cite{Keilson1979}. For a bounded measurable $u$ put
\begin{equation*}
K^*_u=\sup_{0<\lambda<\infty}\ \lambda\int_0^\infty e^{u(t)-\lambda t}\,dt,\qquad
D_{\mathrm{CM}}(f)=\sup_u\Big\{\mathbb{E}_f[u]-\log K^*_u\Big\}. \tag{D.1}
\end{equation*}
By the Donsker--Varadhan formula \cite{DonskerVaradhan1975} and a minimax argument over the compact set of sub-probability measures $\nu$ on $[0,\infty]$, $D_{\mathrm{CM}}(f)=\inf_{g}\mathrm{KL}(f\|g)$ over the vague closure of the completely monotone densities. Hence $D_{\mathrm{CM}}(f)>0$ if and only if $f$ is not completely monotone. The primal problem is convex in $\nu$. Every single test function $u$ already gives a rigorous lower bound $\mathbb{E}_f[u]-\log K^*_u$.

\resultlabel{Lemma D.0}{reversible companion}
Let $Q$ be a finite generator with stationary law $\pi$ and $\sigma(Q)<\infty$. Define $\bar Q$ by $\pi_i\bar Q_{ij}=\bar F_{ij}=\tfrac12(F_{ij}+F_{ji})$ for $i\ne j$. Then $\bar Q$ satisfies detailed balance with respect to $\pi$ and has the same edge set as $Q$. For the stationary path measures on every window $[0,T]$,
\begin{equation*}
\mathrm{KL}\big(\mathbb{P}^{Q}_{[0,T]}\,\big\|\,\mathbb{P}^{\bar Q}_{[0,T]}\big)
=T\sum_{i<j}(F_{ij}+F_{ji})\,\kappa_{\mathrm B}\!\Big(\tfrac{F_{ij}}{F_{ij}+F_{ji}}\Big)\le\frac T4\,\sigma(Q), \tag{D.2}
\end{equation*}
where $\kappa_{\mathrm B}(p)=p\log2p+(1-p)\log2(1-p)$. The constant $1/4$ is sharp near equilibrium.

\begin{proof}
Detailed balance holds by construction, so $\pi$ is stationary for $\bar Q$. Both processes start from $\pi$, so Girsanov's formula for jump processes gives $T^{-1}\mathrm{KL}=\sum_{i\ne j}[F_{ij}\log(F_{ij}/\bar F_{ij})-F_{ij}+\bar F_{ij}]$. The linear terms cancel pairwise, which yields the middle expression. With $p=x/(x+y)$ and $y'=2p-1$,
\[
\kappa_{\mathrm B}(p)=\sum_{k\ge1}\frac{y'^{2k}}{2k(2k-1)},\qquad
\frac{J(x,y)}{4(x+y)}=\frac{y'}4\log\frac{1+y'}{1-y'}=\sum_{k\ge1}\frac{y'^{2k}}{2(2k-1)}.
\]
The series compare termwise, with equality only at $k=1$.
\end{proof}

The geometric-mean reversibilization \cite{DiaconisMiclo2009} gives the weaker rate $\sigma/2-\sum_{i<j}(\sqrt{F_{ij}}-\sqrt{F_{ji}})^2$. In total variation it gives the linear estimate $1-e^{-T\sigma/4}$ used in \cite{AznagulovTwoPrices}. Relative entropy, however, has the advantage of tensorizing over many sojourns, which is exploited next.

\resultlabel{Theorem D}{architecture-independent sojourn floor}
Every finite irreducible realization $(Q,h)$ of $P$, of any dimension and with arbitrary rates, satisfies
\begin{equation*}
\sigma(Q)\ \ge\ 4\,r_C\,D_{\mathrm{CM}}(f_C). \tag{D.3}
\end{equation*}
Consequently $V_n(P)\ge V_\infty(P)\ge4r_CD_{\mathrm{CM}}(f_C)$ for all $n$, and $V_\infty(P)>0$ whenever some sojourn density $f_C$ is not completely monotone. The bound holds in the unrestricted class and in every subclass, including the reset class of Corollary~5.1.

\begin{proof}
Let $B=h^{-1}(C)$ and $A$ its complement among hidden states. Let $\bar Q$ be the companion of Lemma~D.0 and $R$ its observed law. Write $D_A=\mathrm{diag}(\pi_A)$, $D_B=\mathrm{diag}(\pi_B)$. All blocks below are blocks of $\bar Q$.

\emph{Cycle operators.} For bounded $u$ set
\[
E_u=\int_0^\infty e^{u(t)}\,\bar Q_{AB}\,e^{\bar Q_{BB}t}\,\bar Q_{BA}\,dt,\qquad G=(-\bar Q_{AA})^{-1}.
\]
Starting from the hidden state just after an entrance into $A$, the matrix $GE_u$ propagates through the rest of the $A$-stay, one complete $C$-sojourn of duration $\tau$ weighted by $e^{u(\tau)}$, and the next entrance into $A$. Hence $\mathbb{E}_R[\prod_{j\le k}e^{u(\tau_j)}\mid z]=[(GE_u)^k\mathbf1](z)$, and $GE_0$ is stochastic.

\emph{Loewner domination.} Detailed balance gives $D_A\bar Q_{AB}=(D_B\bar Q_{BA})^{\top}$ and makes the matrix $D_B^{1/2}\bar Q_{BB}D_B^{-1/2}=-\sum_k\lambda_kv_kv_k^{\top}$ symmetric, with $\lambda_k>0$ by irreducibility. Therefore
\[
D_AE_u=\sum_k\phi_u(\lambda_k)\,w_kw_k^{\top},\qquad w_k=\bar Q_{BA}^{\top}D_B^{1/2}v_k,\qquad
\phi_u(\lambda)=\int_0^\infty e^{u(t)-\lambda t}dt\le\frac{K^*_u}{\lambda}=K^*_u\,\phi_0(\lambda),
\]
so $0\preceq D_AE_u\preceq K^*_uD_AE_0$. Moreover $S_1=D_A^{1/2}GD_A^{-1/2}$ is symmetric positive definite, and $GE_u$ is similar to $S_1^{1/2}S_2(u)S_1^{1/2}$ with $S_2(u)=D_A^{-1/2}(D_AE_u)D_A^{-1/2}$. This matrix is dominated by $K^*_u$ times $S_1^{1/2}S_2(0)S_1^{1/2}$, which is similar to the stochastic $GE_0$. Hence $\rho(GE_u)\le K^*_u$.

\emph{Supermartingale.} Normalize so that $K^*_u\le1$. By Perron--Frobenius, after an arbitrarily small positive perturbation if $GE_u$ is reducible, there is $\eta>0$ with $GE_u\eta\le\eta$. Then $M_k=\eta(Z_k)\prod_{j\le k}e^{u(\tau_j)}$, with $Z_k$ the hidden state at the $k$-th entrance into $A$, is a positive $R$-supermartingale. Let $G_T$ be the sum of $u(\tau_j)$ over $C$-sojourns that start and end in $[0,T]$, and $N_T$ their number. The index $N_T+1$ is a stopping time, so optional stopping gives $\mathbb{E}_Re^{G_T}\le e^{\|u\|_\infty}\max\eta/\min\eta=:c_R(u)$, independently of $T$.

\emph{Donsker--Varadhan.} By (D.2), data processing through $h$, and the variational formula,
\[
\tfrac T4\,\sigma(Q)\ \ge\ \mathrm{KL}(P_{[0,T]}\|R_{[0,T]})\ \ge\ \mathbb{E}_PG_T-\log\mathbb{E}_Re^{G_T}\ \ge\ \mathbb{E}_PG_T-\log c_R(u).
\]
By the Palm--Campbell formula, the expected sum of $u$ over sojourns starting in $[0,T]$ is $r_CT\,\mathbb{E}_{f_C}u$. At most one of them is incomplete at $T$, so $\mathbb{E}_PG_T\ge r_CT\,\mathbb{E}_{f_C}u-\|u\|_\infty$. Divide by $T$ and let $T\to\infty$. Then replace $u$ by $u-\log K^*_u$ and take the supremum over $u$.
\end{proof}

Only the Palm law of a single observed set enters. No renewal structure, rate bound, dimension bound or reset state is assumed. For an \emph{exactly} reversible realization, Theorem~D reduces to the classical statement behind (9.2). For $P_q$, the observed process is time-reversal invariant, so the standard bound by observed path irreversibility is zero, and (D.3) is the operative bound.

\resultlabel{Corollary D.1}{positive all-architecture floor for the renewal family}
For every $0<q<1$, in the unrestricted and in the reset class,
\begin{equation*}
V_\infty(P_q)\ \ge\ \tfrac43(1+q)\,D_{\mathrm{CM}}(f_q)\ >\ 0. \tag{D.4}
\end{equation*}
For $q=1/20$, the dual certificate (D.1) evaluated at $u=\log(f_q/g^\star)$, with $g^\star$ the numerical primal optimizer, gives $D_{\mathrm{CM}}(f_{1/20})\ge0.07327$, hence $V_\infty(P_{1/20})\ge0.1025$. In particular the conjecture $V_\infty(P_{1/20})=0$ of v1 is false, and scenario~(iii) of \S\ref{sec:scen} does not occur for any $P_q$.

\begin{proof}
By (9.3), $f_q$ is a two-term exponential sum with $B_q<0$. By uniqueness of the Bernstein representation it is not completely monotone, so $D_{\mathrm{CM}}(f_q)>0$. Apply Theorem~D with $C=\{B\}$ and $r=s/3$. The numerical certificate is described in \S\ref{sec:verif}.
\end{proof}

\resultlabel{Proposition 10}{an explicit observable certificate for the zero-price limit}
For every $\kappa>0$ and every bounded $u$,
\[
G_\kappa(P)\ \ge\ V_\infty(P)\ \ge\ 4\,r_C\big(\mathbb{E}_{f_C}[u]-\log K^*_u\big).
\]
The right-hand side depends on the target law only through two observable expectations, the sojourn rate $r_C$ and $\mathbb{E}_{f_C}[u]$. If simultaneous estimates satisfy $|\hat r_C-r_C|\le\epsilon_0$ and $|\hat e-\mathbb{E}_{f_C}u|\le\epsilon_1$, the bound transfers with the obvious error terms, exactly as in (B.6).

\begin{proof}
Combine Theorem~C(4), $G_\kappa\ge V_\infty$, with Theorem~D for a fixed $u$.
\end{proof}

In contrast with the certificates $B_{\kappa,L,m}$ of Theorem~B, whose global calibration is a substantive open task, this certificate is calibrated by construction. It is not tight in general. It is the first explicit lower piece of the architecture phase diagram.

\paragraph{Comparison of floors.} Table~1 compares (D.4) with the optimized three-point total-variation witness of \cite{AznagulovTwoPrices} and with the best explicit realizations of \S\ref{sec:comp}. Far from equilibrium the relative-entropy floor is stronger: by a factor $3.4$ at $q=1/20$ and $4.1$ at $q=0.01$. Near equilibrium the total-variation route wins. The non-monotone amplitude is $|B_q|=\tfrac9{16}(1-q)^2+O((1-q)^3)$. The floor (D.4) scales as $B_q^2\propto(1-q)^4$, whereas the total-variation route gives the sharp order $(1-q)^2$ \cite{AznagulovTwoPrices}. The two bounds are complementary, and $V_\infty(P_q)$ is at least their maximum.

\begin{table}[t]
\centering
\small
\begin{tabular}{@{}ccccc@{}}
\toprule
$q$ & floor (D.4) & three-point TV witness \cite{AznagulovTwoPrices} & best explicit realization & ratio upper/lower\\
\midrule
0.01 & $0.1417$ & $0.0342$ & $4.320^{\,s}$ & 30\\
0.05 & $\mathbf{0.1025}^{\,c}$ & $0.0304$ & $\mathbf{2.7361}^{\,e}$ & 27\\
0.10 & $0.0698$ & $0.0261$ & $2.028^{\,s}$ & 29\\
0.20 & $0.0326^{\,c}$ & $0.0189$ & $1.276^{\,s}$ & 39\\
0.30 & $0.0148$ & $0.0134$ & $0.8397^{\,s}$ & 57\\
0.50 & $0.00248^{\,c}$ & $0.00594$ & $0.3465^{\,m}$ & 58\\
0.70 & $2.2\times10^{-4}$ & $0.00189$ & $0.1070^{\,m}$ & 57\\
0.80 & $3.7\times10^{-5}$ & $7.9\times10^{-4}$ & $0.0446^{\,m}$ & 56\\
\bottomrule
\end{tabular}
\caption{Two-sided bounds on $V_\infty(P_q)$ in units of $k_B$ per unit time. Floors are from the primal convex program; $c$ marks values also certified by the dual (D.1) to relative accuracy $3\times10^{-4}$. Upper bounds: $e$, exact rational realization (Proposition~11); $m$, the three-state machine with $\sigma=(1-q)\log(1/q)$; $s$, special-branch realizations matched to solver precision. The ratio uses the larger of the two lower bounds.}
\end{table}

\section{A rigorous route from these results to useful computation}\label{sec:comp}

The present theory suggests a finite, reviewable objective for future computation. Fix $\kappa>0$, exhibit any exact finite competitor of priced cost $C$, and restrict the search to $n\le C/\kappa$. There is now a proven finite architecture ceiling and an attained optimum. For each permitted dimension, use exact-law equations and globally valid lower certificates. Rate singularities still require suitable coordinates or verified limiting strata; the existence theorem is not a ready-made numerical compact box.

This is a more controlled starting point than asking one computation to certify every dimension at zero state price. It also provides a way to report useful progress: a proven segment of the price-versus-architecture curve, with explicit generators from above and a global lower certificate from below.

\paragraph{The special branch as a convex program.} On the branch $p(S)\mathbf1=0$ of Theorem~6, every phase has an exit density in the span of $e^{-\ell_qu}$ and $e^{-H_qu}$ with total mass one, so positivity forces $0\le t_i\le H_q$. For a fixed vector of exits $t$, the variables $(\alpha,\beta,L)$ enter only through the linear constraints
\[
\textstyle\sum_i\alpha_i=1,\quad \alpha t=b,\quad \alpha_i+\sum_{j}L_{ji}=\sum_jL_{ij}+\beta_it_i,\quad \sum_{j\ne i}L_{ij}(t_j-t_i)=\beta_i(t_i^2-dt_i+c),
\]
and the cost (6.3) is jointly convex. One conic solve per exit vector therefore returns the special-branch optimum, and an outer search over $t$ or a dense grid of exits explores many-phase architectures. A rational exit vector with a numerically optimal flux pattern can be turned into an exact realization: the linear constraints are solved in rational arithmetic by a weighted least-norm correction, Theorem~6 is checked symbolically, and $\sigma$ is evaluated in interval arithmetic.

\resultlabel{Proposition 11}{exact multi-phase improvements of the three-state machine}
For $q=1/20$, the reset class contains exact rational realizations of $P_{1/20}$ with $N=4,5,6$ phases (all edges, entrances and exits strictly positive), whose entropy production satisfies
\[
\sigma_{N=4}<2.760419,\qquad \sigma_{N=5}<2.739512,\qquad \sigma_{N=6}<2.736108.
\]
Their exit vectors are $(\tfrac{97}{2000},\tfrac{77}{100},\tfrac{107}{100},\tfrac65)$, $(\tfrac{97}{2000},\tfrac{13}{20},\tfrac{17}{20},1,\tfrac{57}{50})$ and $(\tfrac{97}{2000},\tfrac{16}{25},\tfrac{41}{50},\tfrac{93}{100},\tfrac{103}{100},\tfrac{57}{50})$. The full rational data are in the ancillary files. Consequently:
\begin{enumerate}[label=(\alph*)]
\item The three-state machine $S=\begin{pmatrix}-s&1\\ q&-s\end{pmatrix}$, $\alpha=(1/s,q/s)$, which realizes $P_q$ with $\sigma=(1-q)\log(1/q)$ ($=2.845946\ldots$ at $q=1/20$), is not an all-dimension optimizer, and $V_7(P_{1/20})<2.736108$.
\item For every $\kappa>0$,
\[
0.1025\ \le\ G_\kappa(P_{1/20})\ \le\ \min\{2.845946+3\kappa,\ 2.760419+5\kappa,\ 2.739512+6\kappa,\ 2.736108+7\kappa\}.
\]
This is the first proven two-sided segment of the architecture phase diagram for $P_{1/20}$.
\end{enumerate}

\begin{proof}
Exactness is verified by checking (6.2) and the transfer function $\alpha(zI-S)^{-1}t=\hat f_{1/20}(z)$ in rational arithmetic, together with $\beta=\alpha(-S)^{-1}$. The entropy production (6.3) is enclosed in 40-digit interval arithmetic. Positivity of all rates gives irreducibility. The lower bound in (b) is Corollary~D.1 and Theorem~C(4).
\end{proof}

In our searches the special-branch optimum did not improve when passing from two to three phases, and improved at four, five and six phases. We do not know whether the sequence continues to decrease. Note that the special branch is not exhaustive (Proposition~7). The gap between $0.1025$ and $2.7361$ is the quantitative form of the open problem: either the lower certificate is loose, or substantially cheaper architectures exist off the special branch.

The state price is a regularization and resource-accounting choice. To interpret it as a hardware cost, a physical implementation model must justify the coefficient. The theorem remains mathematically useful without that interpretation.

\section{Technical proofs and the exact renewal-law case}\label{sec:tech}

The remaining sections carry the weight of the paper: the compactification argument sketched above (\S\S\ref{ss:trace}--\ref{ss:noact}), the exact algebra of the specific renewal law $P_q$ (\S\S\ref{ss:eqs}--\ref{ss:ph4}), and the precise statement of what is still unproved about it (\S\ref{ss:remain} onward). Readers mainly interested in the architecture-price picture of \S\S\ref{sec:floor} and earlier can treat this section as a reference.

\subsection{What complexity divergence actually requires}\label{ss:div}

For finite $V_\infty(P)$, define
\[
C(P,\varepsilon)=\min\{n:\ V_n(P)\le V_\infty(P)+\varepsilon\}.
\]

\resultlabel{Proposition 1}{the exact logical criterion}
\[
C(P,\varepsilon)\longrightarrow\infty \iff V_n(P)>V_\infty(P) \text{ for every finite } n.
\]

\begin{proof}
If $V_n=V_\infty$, then $C(P,\varepsilon)\le n$ for every positive $\varepsilon$. Conversely, if $g_n=V_n-V_\infty>0$, then $\varepsilon<g_n$ excludes every state count at most $n$, because the feasible classes are nested. Since $n$ was arbitrary, the complexity diverges. The set defining $C$ is nonempty for every positive $\varepsilon$ by the definition of the infimum.
\end{proof}

Pointwise nonattainment by each individual finite generator is insufficient on its own: the elementary family $c(n,u)=1+1/u$, $u\ge1$, has no optimizer in any class, yet every fixed-$n$ infimum equals the all-class infimum $1$. A parameter escaping to infinity can eliminate a dimension gap even though no finite object ever achieves it.

Nor is the following statement valid without restrictions on the observed law: ``Every finite irreversible realization can be strictly improved by adding hidden states.''

\resultlabel{Proposition 2}{a counterexample to universal irreversible improvement}
Let the observed process itself be the three-state ring with clockwise rate $a>0$, counterclockwise rate $b>0$, and $a\neq b$. Its finite realization is globally optimal among all finite deterministic hidden realizations of that full observed law, and
\[
V_\infty(P) = (a-b)\log(a/b) > 0.
\]

\begin{proof}
The uniform stationary law gives the displayed entropy production. For every hidden realization, path-space relative entropy decreases under observation. For a stationary Markov jump process, path divergence from its time reversal over a time interval of length $T$ equals $T\sigma$ \citep{Seifert2012}. The observed process is the specified Markov chain, so $T\sigma(Q)\ge T(a-b)\log(a/b)$. Identity observation attains this bound.
\end{proof}

This counterexample does not settle the conjecture for the reversible observed renewal law $P_q$. It does show that irreversibility of the hidden model alone cannot establish a universal improvement theorem. A successful $P_q$-specific proof must use more than nonzero entropy production.

\subsection{Trace contraction of entropy production}\label{ss:trace}

Let $A$ be a nonempty subset of the states of an irreducible finite chain, and $B=A^c$. For $B\neq\varnothing$, the trace generator on $A$ is
\[
Q^A = Q_{AA} + Q_{AB}(-Q_{BB})^{-1}Q_{BA}.
\]
It is obtained by deleting the time spent in $B$. Its stationary law is
\[
\pi^A_i = \frac{\pi_i}{\pi(A)}, \qquad i\in A.
\]
These are the continuous-time trace or stochastic-complement formulas. In an irreducible finite chain, $B$ is transient for the chain killed on hitting $A$, so the inverse exists. If $A$ is the full state set, all statements below reduce to identities.

\resultlabel{Theorem 3}{entropy contraction with the correct clock}
\begin{equation*}
\pi(A)\,\sigma(Q^A) \le \sigma(Q). \tag{3.1}
\end{equation*}
Moreover,
\begin{equation*}
\pi(A)\,K(Q^A) \le K(Q), \qquad \pi_i \sum_{j\in A\setminus\{i\}} Q^A_{ij} \le K(Q). \tag{3.2}
\end{equation*}
The factor $\pi(A)$ converts trace time to original stationary time. Omitting it would in general assert the wrong inequality.

\begin{proof}
Infinite original entropy production makes (3.1) immediate, so assume finite cost; every positive edge then has its reverse. Consider all paths
\[
\gamma=(i,b_1,\dots,b_m,j), \qquad i,j\in A,\ b_\ell\in B,
\]
with no intermediate visit to $A$, allowing repeated $B$ states. Include direct transitions $i\to j$, and include return excursions $i=j$ when $m\ge1$. Their stationary occurrence intensities, per unit original time, are
\[
w_\gamma = \pi_i Q_{ib_1}\left(\prod_{\ell=1}^{m-1}\frac{Q_{b_\ell b_{\ell+1}}}{\lambda_{b_\ell}}\right)\frac{Q_{b_m j}}{\lambda_{b_m}}, \qquad \lambda_x=-Q_{xx},
\]
and $w_{(i,j)}=\pi_i Q_{ij}$ for direct edges. The denominator product is unchanged by reversal. Therefore
\begin{equation*}
\log\frac{w_\gamma}{w_{\gamma^R}} = \sum_{(u,v)\text{ along }\gamma} \log\frac{\pi_u Q_{uv}}{\pi_v Q_{vu}}. \tag{3.3}
\end{equation*}
Every stationary jump belongs to exactly one such excursion or direct transition. Expected edge traversal counts per unit original time are $F_{uv}$. The killed finite chain has finite mean return time, so this counting also justifies summing the absolute values of the finitely many edge log ratios. Consequently
\begin{equation*}
\sum_\gamma w_\gamma \log\frac{w_\gamma}{w_{\gamma^R}} = \sigma(Q). \tag{3.4}
\end{equation*}
For distinct $i,j\in A$, sum the excursion intensities with these endpoints:
\[
W_{ij} = \sum_{\gamma:\,i\to j} w_\gamma = \pi_i Q^A_{ij}.
\]
The log-sum inequality in each endpoint class bounds its contribution below by $W_{ij}\log(W_{ij}/W_{ji})$. A return class $i=j$ has the same total intensity as its reversal and contributes at least zero. Summing yields (3.1), because the $W_{ij}$ use the original, unnormalized mass on $A$.

An excursion from $i$ ending at a different state of $A$ occurs no more often than departures from $i$. Thus
\[
\sum_{j\in A\setminus\{i\}} W_{ij} \le \pi_i\lambda_i.
\]
Summing gives (3.2).
\end{proof}

\subsection{Compactification under bounded mean activity}

Here is a finite-dimensional closure theorem that tolerates individual rates diverging to infinity. It uses bounded mean activity, a weaker hypothesis than a uniform rate cap, and it is the base case for the fully general Theorem~5 below.

\resultlabel{Theorem 4}{bounded-activity exact-law compactification}
Let $P$ be an ergodic stationary law on a finite alphabet with c\`{a}dl\`{a}g paths. Suppose $(Q_k,\pi^{(k)},h_k)$ are finite irreducible stationary realizations of exactly $P$, with
\[
|Q_k|\le n, \qquad \sup_k K(Q_k)<\infty, \qquad L:=\liminf_k \sigma(Q_k)<\infty.
\]
Then a finite irreducible stationary realization $Q^*$ of exactly $P$ exists with
\begin{equation*}
|Q^*|\le n, \qquad \sigma(Q^*)\le L. \tag{4.1}
\end{equation*}
No bound on $\max_i(-Q_{k,ii})$ or positive lower bound on the entries of $\pi^{(k)}$ is assumed. This is a theorem about the unrestricted realization class. It need not preserve an imposed graph topology or an additional kinetic constraint. Preservation of a single-reset-state architecture is proved separately in Corollary~5.1.

\begin{proof}
Pass to a subsequence achieving the $\liminf$ of costs. There are finitely many state counts and observation maps, up to relabeling, so fix both. Compactness of the probability simplex gives $\pi^{(k)}\to\pi^0$. Let
\[
A=\{i:\pi^0_i>0\}, \qquad B=A^c, \qquad \delta_k=\pi^{(k)}(B)\longrightarrow0.
\]
Write $R_k=Q^A_k$. By (3.2), for $i\in A$,
\begin{equation*}
\sum_{j\in A\setminus\{i\}} (R_k)_{ij} \le \frac{K(Q_k)}{\pi^{(k)}_i}. \tag{4.2}
\end{equation*}
The right side is uniformly bounded for large $k$. Thus a further subsequence satisfies $R_k\to R_0$, a finite generator on $A$, and $\pi^{(k),A}\to\pi^0_A$, whose entries are all positive and sum to one. Stationarity passes to the limit.

We next verify that deleting the disappearing states preserves the observed law in the limit. Start the original process conditional on $X_0\in A$. Its initial law differs in total variation from the stationary law by $\delta_k$. Define
\[
C_k(u) = \int_0^u \mathbf{1}\{X_s\in A\}\,ds, \qquad \tau_k(t) = \inf\{u:\ C_k(u)>t\}.
\]
The trace $X_{\tau_k(t)}$ is stationary for $R_k$. Let $H_k(U)$ be time spent in $B$ up to $U$. Conditioning costs at most the factor $1/(1-\delta_k)$, so
\[
\mathbb{E}[H_k(U)\mid X_0\in A] \le \frac{U\delta_k}{1-\delta_k}.
\]
Since $\tau_k(T)-T>\eta$ implies $H_k(T+\eta)\ge\eta$,
\begin{equation*}
\Pr(\tau_k(T)-T>\eta \mid X_0\in A) \le \frac{(T+\eta)\delta_k}{(1-\delta_k)\eta}. \tag{4.3}
\end{equation*}
For finitely many times $t_1,\dots,t_m\le T$, outside this event the time shifts all lie in $[0,\eta]$. Observations agree if no observed jump occurs in any $[t_j,t_j+\eta]$. Under the common stationary law $P$, that probability of a jump tends to zero as $\eta\downarrow0$: a finite-alphabet c\`{a}dl\`{a}g path has finitely many jumps on a compact interval, and stationarity excludes a jump at a prescribed deterministic time. The conditional probability is at most its unconditional value divided by $1-\delta_k$. First send $k\to\infty$ in (4.3), then $\eta\downarrow0$. The observed finite-dimensional laws of the stationary traces therefore converge to those of $P$.

Matrix exponentials are continuous in a finite generator. Hence $(R_0,\pi^0_A,h|_A)$ has all the finite-dimensional distributions of $P$. Both laws are c\`{a}dl\`{a}g; equality on all finite-dimensional cylinders determines their path laws.

The extended nonnegative function $J$ is lower semicontinuous on the nonnegative quadrant. Theorem 3 gives
\begin{equation*}
\sigma(R_0,\pi^0_A) \le \liminf_k \sigma(R_k,\pi^{(k),A}) \le L, \tag{4.4}
\end{equation*}
because $\pi^{(k)}(A)\to1$ and $\sigma(Q_k)\to L$.

Finally $R_0$ may be reducible. A strictly positive stationary law on a finite generator places mass only on its closed communicating classes, so it is a finite stationary mixture of irreducible classes. The observed component laws are stationary and average to the ergodic law $P$. Extremality of an ergodic law forces each positive-weight component law to equal $P$. Entropy production is the weighted average of component costs; select a component whose cost is at most (4.4). It has at most $n$ states and proves (4.1).
\end{proof}

The next result removes the activity assumption altogether.

\subsection{Removing the activity bound}\label{ss:noact}

\resultlabel{Theorem 5}{fixed-state-count compactification and attainment}
Let $P$ be an ergodic stationary finite-alphabet c\`{a}dl\`{a}g law. Let $Q_k$ be finite irreducible stationary exact realizations of $P$, with at most $n$ states, and let
\[
L=\liminf_k \sigma(Q_k) < \infty.
\]
Then a finite irreducible exact realization $Q^*$ exists with at most $n$ states and $\sigma(Q^*)\le L$. No rate bound, mean-activity bound, or prescribed hierarchy of time scales is required. In particular, every finite $V_n(P)$ is attained.

\begin{proof}[Proof, step 1: remove disappearing occupation]
Pass to a subsequence attaining the $\liminf$ and fix the state count and observation map. Let $\pi^{(k)}\to\pi^0$, and trace on the states with positive limiting mass, exactly as in Theorem 4. Crucially, the time-change estimate (4.3) and the preservation of observed finite-dimensional laws did not use bounded activity. The trace laws converge to $P$, their stationary probabilities are uniformly bounded away from zero, and
\begin{equation*}
\limsup_k \sigma(R_k) \le L. \tag{5.1}
\end{equation*}
It remains to compactify these possibly unbounded trace generators $R_k$. Rename them $Q_k$ and their stationary distributions $\pi^{(k)}$. There are now $m\le n$ states with $\pi^{(k)}_i\to\pi^0_i>0$. Their observed laws converge to $P$, rather than necessarily equaling $P$ at finite $k$.
\end{proof}

\begin{proof}[Step 2: identify all divergent conductances at once]
Put
\[
c^{(k)}_{ij} = \tfrac12\left(\pi^{(k)}_i Q_{k,ij} + \pi^{(k)}_j Q_{k,ji}\right), \qquad i<j.
\]
Because there are finitely many pairs, pass to a subsequence along which every $c^{(k)}_{ij}$ converges in $[0,\infty]$. Connect $i,j$ when their conductance tends to infinity, and let $C_1,\dots,C_r$ be the connected components of this undirected graph, including isolated vertices. All fluxes between different components are bounded. Pass again to a subsequence where each such oriented flux converges.

Let $w^{(k)}_a=\sum_{i\in C_a}\pi^{(k)}_i$, and let $\Pi_k$ be the conditional-expectation projection onto block-constant functions in $L^2(\pi^{(k)})$:
\[
(\Pi_k f)_i = \frac{1}{w^{(k)}_a}\sum_{j\in C_a} \pi^{(k)}_j f_j, \qquad i\in C_a.
\]
Let $E_k=I-\Pi_k$. Both projections have norm at most one.
\end{proof}

\begin{proof}[Step 3: fast coercivity without reversibility]
Stationarity alone gives the real Dirichlet identity
\begin{equation*}
-\langle f,Q_kf\rangle_{\pi^{(k)}} = \sum_{i<j} c^{(k)}_{ij}(f_i-f_j)^2. \tag{5.2}
\end{equation*}
Choose a spanning tree of divergent-conductance edges in each nonsingleton block. Their minimum conductance tends to infinity. The finite graph Poincar\'{e} inequality, with the stationary probabilities uniformly bounded away from zero, supplies $\lambda_k\to\infty$ such that
\begin{equation*}
-\langle v,Q_kv\rangle_{\pi^{(k)}} \ge \lambda_k \|v\|^2_{\pi^{(k)}}, \qquad \Pi_kv=0. \tag{5.3}
\end{equation*}
For clarity, the required Poincar\'{e} inequality is elementary: on a fixed tree, every difference $v_i-v_j$ is a sum of at most $m-1$ edge differences; square and sum, and use the zero weighted block means. This bounds the weighted variance by a constant times the tree-edge squared differences, uniformly in $k$. If there are only singleton blocks, every flux and every rate is bounded and the earlier finite-generator limit argument already finishes the proof.
\end{proof}

\begin{proof}[Step 4: bound both slow-fast couplings]
Define the four compressed operators
\[
\mathcal{A}_k=\Pi_kQ_k\Pi_k, \qquad \mathcal{B}_k=E_kQ_k\Pi_k, \qquad \mathcal{C}_k=\Pi_kQ_kE_k, \qquad \mathcal{D}_k=E_kQ_kE_k.
\]
The operators $\mathcal{A}_k,\mathcal{B}_k,\mathcal{C}_k$ have uniformly bounded norms. For a block-constant vector $f$, internal transitions cancel in $Q_k\Pi_kf$ and only cross-block rates remain. Every cross-block oriented stationary flux is bounded, and every stationary probability is bounded below after Step 1, so all such rates are uniformly bounded; this controls $\mathcal{A}_k$ and $\mathcal{B}_k$. For $\mathcal{C}_k$, use the adjoint in $L^2(\pi^{(k)})$,
\[
(Q_k^\dagger)_{ij}=\frac{\pi^{(k)}_jQ_{k,ji}}{\pi^{(k)}_i}\quad(i\neq j).
\]
Its cross-block rates are bounded by the reverse cross-block fluxes divided by the same positive stationary masses. Since $\mathcal{C}_k$ is the adjoint of $E_kQ_k^\dagger\Pi_k$, it is uniformly bounded as well. This is the point that rules out a large antisymmetric slow-fast coupling hidden behind the diverging symmetric conductances.

For $v$ in the $E_k$ subspace, $E_kv=v$ and orthogonality of $\Pi_k$ gives
\[
\operatorname{Re}\langle v,\mathcal{D}_kv\rangle_{\pi^{(k)}}
=\operatorname{Re}\langle v,Q_kv\rangle_{\pi^{(k)}}
\le-\lambda_k\|v\|^2_{\pi^{(k)}}.
\]
Hence, differentiating $\|e^{t\mathcal D_k}v\|^2$ and applying Gr\"onwall,
\[
\|e^{t\mathcal D_k}\|_{E_k\to E_k}\le e^{-\lambda_kt}.
\]
The full Markov semigroup is an $L^2(\pi^{(k)})$ contraction by Jensen's inequality and stationarity. On the block-constant subspace, $\mathcal A_k$ is the generator of the block chain in (5.6), with stationary law $w^{(k)}$, and its semigroup is likewise contractive.
\end{proof}

\begin{proof}[Step 5: an explicit semigroup estimate]
For $f_k(t)=e^{tQ_k}f$, write $u_k=\Pi_kf_k$ and $v_k=E_kf_k$. Then
\[
u_k' = \mathcal{A}_ku_k+\mathcal{C}_kv_k, \qquad v_k' = \mathcal{B}_ku_k+\mathcal{D}_kv_k.
\]
Let $M$ bound $\|\mathcal B_k\|$ and $\|\mathcal C_k\|$. Since $\|u_k(t)\|\le\|f\|$, variation of constants in the fast equation gives
\[
\|v_k(t)\|\le e^{-\lambda_kt}\|v_k(0)\|+
M\int_0^t e^{-\lambda_k(t-s)}\|u_k(s)\|\,ds
\le e^{-\lambda_kt}\|f\|+\frac{M}{\lambda_k}\|f\|.
\]
Substituting this in the slow equation yields
\begin{align*}
\|u_k(t)-e^{t\mathcal{A}_k}\Pi_kf\|
&\le M\int_0^t\|v_k(s)\|\,ds\\
&\le \frac{M+M^2t}{\lambda_k}\|f\|. \tag{5.4}
\end{align*}
Therefore, for every $0<\eta<T<\infty$,
\begin{equation*}
\sup_{t\in[\eta,T]}
\left\|e^{tQ_k}-He^{t\overline{Q}_k}W^k\right\|\longrightarrow0. \tag{5.5}
\end{equation*}
Here $H$ lifts block functions to state functions, $W^k$ takes stationary block averages, $\Pi_k=HW^k$, and
\begin{equation*}
(\overline{Q}_k)_{ab} = \frac{1}{w^{(k)}_a}\sum_{i\in C_a,\,j\in C_b} \pi^{(k)}_iQ_{k,ij}, \qquad a\neq b. \tag{5.6}
\end{equation*}
The cross-block flux bounds make $\overline Q_k$ uniformly bounded, so after a subsequence it converges to a finite generator $\overline Q$, stationary for $w_a=\sum_{i\in C_a}\pi^0_i>0$. Because the stationary probabilities stay uniformly away from zero, all varying weighted norms above are uniformly equivalent to the Euclidean norm; thus (5.5) is also ordinary finite-matrix convergence.
\end{proof}

\begin{proof}[Step 6: stochastic continuity makes every fast block monochromatic]
For each observation symbol $y$, let
\[
p_a(y)=\frac{1}{w_a}\sum_{i\in C_a:\,h(i)=y}\pi_i^0
\]
be its stationary conditional frequency inside block $a$. The semigroup limit (5.5) gives, for each fixed $t>0$,
\[
\Pr(Y_0=Y_t)\longrightarrow
\sum_{a,b}w_a(e^{t\overline Q})_{ab}\sum_y p_a(y)p_b(y).
\]
(The left side is already the fixed observed probability for the pre-trace sequence, or converges to it after Step 1.) Sending $t\downarrow0$ and using stochastic continuity of the c\`{a}dl\`{a}g target law gives
\begin{equation*}
0 = \lim_{t\downarrow0} \Pr\nolimits_P(Y_0\neq Y_t) = \sum_a w_a\left(1-\sum_y p_a(y)^2\right). \tag{5.7}
\end{equation*}
Every summand is nonnegative and every $w_a>0$, so each $p_a$ is a point mass. Since all limiting microscopic stationary masses are positive, the deterministic observation map is constant on each fast block $C_a$.

For a symbol $y$, the diagonal observation projector is therefore block-constant and commutes with $\Pi_k$ for all sufficiently large $k$ (the state labels were fixed before taking subsequences). Insert (5.5) between these projectors in any finite-dimensional cylinder probability, keeping all positive time gaps fixed. The limit is exactly the corresponding cylinder probability of $(\overline Q,w,h)$. Those cylinder probabilities equal the target ones, and finite-alphabet c\`{a}dl\`{a}g laws are determined by their finite-dimensional distributions. Thus the reduced CTMC realizes the full observed path law.
\end{proof}

\begin{proof}[Step 7: entropy cannot increase]
Discard within-block entropy and apply log-sum to cross-block pairs:
\[
\sigma(Q_k) \ge \sum_{a<b} J\!\left(F^{(k)}_{ab},F^{(k)}_{ba}\right), \qquad F^{(k)}_{ab}=w^{(k)}_a(\overline{Q}_k)_{ab}.
\]
Taking the $\liminf$ gives $\sigma(\overline{Q},w)\le L$, using (5.1). As in Theorem 4, select an irreducible component. Ergodicity of $P$ makes its observed law equal $P$, and some component has no larger entropy production. State count has only decreased. Applying the statement to a minimizing sequence proves attainment of finite $V_n(P)$.
\end{proof}

\resultlabel{Corollary 5.1}{the fixed-$N$ escape problem in the reset PH class}
For an ergodic alternating renewal law with one exponentially distributed reset block $A$ and a positive finite mean $B$ dwell, the finite-cost infimum over at most $N$ $B$-phases is attained within that class.

\begin{proof}
The unique microscopic $A$ state has a fixed positive stationary mass determined by the observed means, so it is never deleted in step 1. Equation (5.7) prevents it from merging with a $B$ state. Thus the limit still has exactly one $A$ state. Every positive-weight recurrent component must generate $P$, which visits $A$; two disjoint such components cannot both contain the unique $A$ state. Hence the used limit is irreducible, and the same law fixes the reset exit rate. At most $N$ $B$-phases remain.
\end{proof}

This includes the fixed law $P_{1/20}$, provided the infimum is finite; its finite positive-rate two-phase realization supplies that finiteness for all $N\ge2$. If exactly $N$ used phases are required, balanced splitting of a $B$ state pads a smaller optimizer without changing law or cost.

The proof is not a numerical global optimization. It establishes existence of an optimizer, not its value, location, uniqueness, or sparsity. Its key tools are classical; the exact-law, entropy-lower-semicontinuous compactification assembled here warrants independent specialist review and a dedicated priority check.

\subsection{Exact all-dimension equations for the degree-two dwell law}\label{ss:eqs}

With the general compactification in hand, we turn to the one renewal law the whole paper is really about, and ask what a finite phase-type representation of it must satisfy exactly, at any finite phase count. Fix $q=1/20$, or more generally $0<q<1$, and set
\[
s=1+q, \qquad d=2s, \qquad c=1+q+q^2, \qquad b=\frac{2q}{1+q}, \qquad p(z)=z^2+dz+c.
\]
The $B$-dwell transform is
\begin{equation*}
\hat f_q(z) = \frac{bz+c}{p(z)}. \tag{6.1}
\end{equation*}
The $A$ dwell is exponential with rate $s$. An $N$-phase representation consists of a transient Metzler matrix $S\in\mathbb{R}^{N\times N}$, $t=-S\mathbf{1}\ge0$, and an entrance row vector $\alpha\ge0$, $\alpha\mathbf{1}=1$. Transient means every eigenvalue has strictly negative real part. The full reset generator is
\[
Q = \begin{pmatrix} -s & s\alpha \\ t & S \end{pmatrix}.
\]
Successive returns to the unique $A$ state guarantee the full alternating renewal law, once (6.1) is matched. Unused states can be deleted.

\resultlabel{Theorem 6}{complete finite cancellation equations}
Such a PH representation has transform (6.1) if and only if
\begin{equation*}
\alpha t = b, \qquad \alpha S^k p(S)\mathbf{1} = 0, \quad k=0,1,\dots,N-1. \tag{6.2}
\end{equation*}
No diagonalizability, simple hidden eigenvalues, or one-sided Kalman-branch assumption is required.

\begin{proof}
The survival function $g(u)=\alpha e^{Su}\mathbf{1}$ is analytic, with $g(0)=1$ and $g'(0)=-\alpha t$. Its residual in the scalar ODE is
\[
g''+dg'+cg = \alpha e^{Su}p(S)\mathbf{1}.
\]
Under (6.2), Cayley--Hamilton makes every power $S^k$ with $k\ge N$ a linear combination of earlier powers, so all Taylor coefficients of this residual vanish. Therefore $g''+dg'+cg=0$, $g(0)=1$, and $g'(0)=-b$. Laplace transformation yields
\[
\hat g(z) = \frac{z+d-b}{p(z)}, \qquad f(z) = 1-z\hat g(z) = \frac{bz+c}{p(z)}.
\]
Conversely, the stated rational transform implies this ODE and initial derivative. Differentiating its residual at zero gives every equality in (6.2).
\end{proof}

The equations are polynomial in $S,\alpha$, but not jointly convex. In particular, (6.2) only says that $p(S)\mathbf{1}$ lies in the unobservable subspace. It does not require $p(S)\mathbf{1}=0$ or $\alpha p(S)=0$ --- a distinction that turns out to matter, as Proposition~7 shows below.

For used phases, let $\beta=\alpha(-S)^{-1}$ and $L_{ij}=\beta_iS_{ij}$. Then $\sum_i\beta_i=2/s$, the entry rate into either observed block is $r=s/3$, and
\begin{equation*}
\sigma(Q) = r\left[\sum_i J(\alpha_i,\beta_it_i) + \sum_{i<j} J(L_{ij},L_{ji})\right]. \tag{6.3}
\end{equation*}
On the special branch $p(S)\mathbf{1}=0$, multiplication by $\beta_i$ gives the previously used affine drift equations
\[
\sum_{j\neq i} L_{ij}(t_j-t_i) = \beta_i(t_i^2-dt_i+c).
\]
Together with flux balance, these give a convex inner problem at fixed exits $t$. Theorem 6 is the correct all-$N$ replacement for declaring that branch globally exhaustive.

\subsection{An exact four-phase obstruction to the two-branch reduction}\label{ss:ph4}

The convex branch just described is tempting to treat as exhaustive. It is not, and the next example shows why concretely rather than abstractly.

\resultlabel{Proposition 7}{the obstruction, made explicit}
At $q=1/20$, the following strictly positive-edge PH representation has precisely transform (6.1), but
\[
p(S)\mathbf{1}\neq0, \qquad \alpha p(S)\neq0.
\]
Take
\[
S=\begin{pmatrix}
-2699/2000 & 601/2000 & 1001/2000 & 999/2000\\
599/2000 & -2701/2000 & 1001/2000 & 999/2000\\
1/40 & 1/40 & -29/20 & 2/5\\
1/40 & 1/40 & 2/5 & -29/20
\end{pmatrix},
\]
\[
\alpha = (10/21,\ 10/21,\ 1/42,\ 1/42), \qquad t = (49/1000,\ 51/1000,\ 1,\ 1)^\top.
\]

\begin{proof}
All off-diagonal rates, entrances, and exits are positive. The row sums equal $-t_i$. Strict row diagonal dominance with negative diagonal ensures transience. To exhibit the exact cancellations, put
\[
U=\begin{pmatrix}
1&0&1&0\\
1&0&-1&0\\
0&1&0&1\\
0&1&0&-1
\end{pmatrix}, \qquad
U^{-1}SU = \begin{pmatrix}
-21/20 & 1 & 0 & 1/1000\\
1/20 & -21/20 & 0 & 0\\
1/1000 & 0 & -33/20 & 0\\
0 & 0 & 0 & -37/20
\end{pmatrix}.
\]
Here $\alpha U=(20/21,1/21,0,0)$ and $U^{-1}\mathbf{1}=(1,1,0,0)^\top$. The fourth coordinate, initially zero, remains zero; the third may be driven but does not feed back into the first two. Thus the survival function is exactly that of the upper-left two-state block, giving (6.1). Direct rational multiplication gives
\[
p(S)\mathbf{1} = (1/2500,\ -1/2500,\ 0,\ 0)^\top, \qquad
\alpha p(S) = (0,\ 0,\ -319/840000,\ 319/840000).
\]
Both controllability and observability ranks are $3$, while the scalar transfer function has reduced degree $2$. Explicitly,
\[
\alpha(zI-S)^{-1}t = \frac{800z+8841}{21(400z^2+840z+421)}.
\]
This is exactly (6.1).
\end{proof}

The example includes coincident exits but neither cancellation identity holds. The coincidence therefore does not restore the two-branch coverage. It does not contradict the previous $N=3$ argument, which used its special dimension. It also makes no assertion about the entropy optimality of this four-phase model.

The associated phenomenon is standard nonminimal realization structure: a mode can be reachable but unobservable while a different mode is observable but unreachable. The contribution here is an explicit positive PH instance for the exact law under investigation.

\subsection{The remaining theorem, stated without hidden assumptions}\label{ss:remain}

Theorem 5 and Corollary 5.1 remove the need for a separate compactness hypothesis in their stated classes. For a fixed ergodic $P$ with finite $V_\infty(P)$, the remaining substantive statement is:

\begin{quote}
\emph{Every finite exact realization of $P$ can be strictly improved by another finite exact realization, whose dimension need not increase by just one.}
\end{quote}

\resultlabel{Theorem 8}{equivalent remaining targets}
In the unrestricted class, the following are equivalent:
\begin{enumerate}[label=(\arabic*)]
\item $V_n(P)>V_\infty(P)$ for every finite $n$.
\item No finite exact realization attains $V_\infty(P)$.
\item Every finite exact realization can be strictly improved by some finite exact realization.
\item $C(P,\varepsilon)\to\infty$.
\end{enumerate}
The same equivalence holds within the single-reset PH class of Corollary 5.1, with its own infima and phase count.

\begin{proof}
Statement 1 implies 2 directly. If 2 holds and a finite realization has cost $c>V_\infty$, the definition of the infimum supplies a finite exact realization of cost less than $c$; the same is true when $c=+\infty$. Thus 2 implies 3. Statement 3 excludes any finite all-class optimizer. If 1 were false, Theorem 5 would attain a fixed-dimension value equal to $V_\infty$, contradicting 3. Proposition 1 gives the equivalence with 4. Corollary 5.1 supplies the identical attainment step in the reset class.
\end{proof}

Theorem 8 is an equivalence, not a proof that any of its statements holds for $P_q$. Its useful advance over a purely formal reduction is that the fixed-dimension attainment step has a proof in Theorem 5. The remaining nonattainment/improvement assertion is false for arbitrary observed laws by Proposition 2, and no valid $P_q$-specific proof is currently supplied.

For $P_q$, every finite realization is irreversible because the complete dwell density has a negative exponential spectral weight, and by Corollary~D.1 the irreversibility is bounded below uniformly in dimension. Neither fact proves statement 2. Proposition~11 shows that statement 3 does hold for the simplest realization, the three-state machine, but the general case is open. A successful next theorem must exploit exact-law hidden cancellations or another property specific to this reversible observed renewal law. Proposition 7 shows why restricting all competitors to the old $N=3$ convex branch would leave a gap.

Nor does strict improvement at a fixed $N$ require an infinitesimal profitable split. An optimizer may resist every local split while admitting a better distant architecture. The appropriate target is a finite exact competitor with strictly smaller cost, without a restriction to one extra pole or one extra phase.

\subsection{A strict finite-dimension consequence for the renewal family}

\resultlabel{Theorem 9}{positive dissipation in every feasible finite dimension}
For every $0<q<1$, every feasible finite fixed-state-count infimum $V_n(P_q)$ is strictly positive. This holds in the unrestricted deterministic-observation class and in the single-reset PH subclass. In particular it holds for $q=1/20$.

\begin{proof}
In a reversible finite chain, let $B$ be one observed block and let $S=Q_{BB}$ be its killed generator. Write $t_i=\sum_{j\notin B}Q_{ij}$ and let $r=\sum_{i\in B}\pi_it_i>0$ be the stationary exit intensity. Detailed balance makes the stationary entrance intensity into $i$ equal to $\pi_it_i$. Consequently the density of a complete $B$ dwell, sampled at an entrance, is
\begin{equation*}
f_B(u) = \frac{1}{r}\langle t,e^{uS}t\rangle_{\pi|B}. \tag{9.1}
\end{equation*}
Detailed balance inside $B$ makes the killed generator self-adjoint in this weighted inner product. Writing its eigenvalues as $-\lambda_\ell<0$ and expanding in an orthonormal eigenbasis,
\begin{equation*}
f_B(u) = \frac{1}{r}\sum_\ell |\langle t,\psi_\ell\rangle_{\pi|B}|^2 e^{-\lambda_\ell u}. \tag{9.2}
\end{equation*}
Thus every complete dwell density generated by a finite reversible realization is a nonnegative finite mixture of decaying exponentials.

For the target family, factor
\[
p(z)=(z+\ell_q)(z+H_q),\qquad
\ell_q=1+q-\sqrt q,\quad H_q=1+q+\sqrt q.
\]
Its dwell transform has the unique partial-fraction decomposition
\begin{equation*}
\widehat f_q(z)=\frac{A_q}{z+\ell_q}+\frac{B_q}{z+H_q}, \qquad
A_q=\frac{(1+q^{3/2})^2}{2\sqrt q(1+q)},\quad
B_q=-\frac{(1-q^{3/2})^2}{2\sqrt q(1+q)}. \tag{9.3}
\end{equation*}
For $0<q<1$, $B_q<0$. Uniqueness of the Laplace transform (equivalently, uniqueness of the residues of this rational function) therefore rules out any representation of $f_q$ as a nonnegative finite exponential mixture. Hence no finite reversible exact realization of $P_q$ exists.

If a feasible finite $V_n(P_q)$ were zero, Theorem 5 would attain it. Zero entropy production in an irreducible finite CTMC forces detailed balance on every edge, contradicting the preceding obstruction. Therefore $V_n(P_q)>0$. The reset-class statement follows from Corollary 5.1 in the same way.
\end{proof}

At $q=1$ the negative residue vanishes and the dwell law collapses to a single exponential, so the obstruction closes exactly at the equilibrium endpoint. For $q=1/20$, one may also see the contradiction locally from $f'_{1/20}(0)=341/400>0$, whereas every mixture in (9.2) has nonpositive derivative at the origin.

The equilibrium-mixture obstruction is consistent with earlier waiting-time results \citep{SkinnerDunkel2021}. The new role of Theorem 5 is to upgrade ``every individual finite model dissipates'' to a \emph{uniform positive infimum at each fixed dimension}, even when microscopic rates are unbounded. Theorem 9 by itself gives no lower bound uniform in $n$. In v1 this left open whether $V_n(P_q)$ approaches a positive floor or tends to zero as hidden architecture grows. Corollary~D.1 now settles this: the floor is positive, and Theorem 9 is a special case of Theorem~D.

\subsection{A dimension-padding construction}\label{ss:pad}

For completeness, the padding assertion used above has an exact construction. Replace one state $i$ by clones $i_1,\dots,i_k$ of the same observed label and choose positive weights $\theta_a$ summing to one. Set
\[
Q'_{j,i_a} = \theta_aQ_{ji}, \qquad Q'_{i_a,j} = Q_{ij}\ (j\neq i), \qquad Q'_{i_a,i_b} = \kappa\theta_b\ (a\neq b)
\]
with any $\kappa>0$, leaving other rates unchanged and setting diagonal entries by row sums. The stationary masses are $\pi'_{i_a}=\theta_a\pi_i$. The clone partition is strongly lumpable to the original generator, so the observed law is unchanged. Clone-internal stationary fluxes are symmetric and contribute zero entropy. Each external pair contributes $\theta_aJ(F_{ij},F_{ji})$; summing over clones reproduces the original cost. Irreducibility is preserved. Splitting a $B$ phase preserves a unique reset state. Thus exact used dimension and an at-most dimension ceiling have the same infimum in these unrestricted classes.

\section{Verification, reproducibility, and relation to prior work}\label{sec:verif}

To our knowledge, the exact-law, activity-unbounded attainment theorem and the resulting state-priced observable-certificate representation have not been stated in this combined form. This is a deliberately narrow priority statement, not a claim that fast-state reduction, coarse-graining inequalities, or hidden-state thermodynamic optimization are themselves new.

Supplementary Code S1, \texttt{verify\_thermodynamic\_realizability.py}, verifies the four-phase identities of Proposition 7 in exact rational arithmetic, including positivity, normalization, the transfer function, both Kalman ranks, and all cancellation conditions. It also checks trace contraction and a multiscale semigroup reduction on small finite chains as floating-point diagnostics. These computations do not verify Theorems 3--5: the compactness and lower-semicontinuity arguments are analytic. No global $N=3$ optimization is used in this paper.

Version~2 adds three scripts. \texttt{floor\_certificate.py} solves the primal convex program for $D_{\mathrm{CM}}(f_q)$ on a grid of 500 rates and evaluates the Donsker--Varadhan dual (D.1) at $u=\log(f_q/g^\star)$ by adaptive quadrature. It evaluates $K^*_u$ on 4000 logarithmically spaced rates in $[10^{-4},10^5]$ together with the limits $\lambda\to0,\infty$. At $q=1/20$ it returns primal $0.073298$, dual $0.073270$ and $K^*_u=1.000028$. Because $u$ is unbounded below at large times, (D.1) is applied to $\max(u,-M)$ and $M\to\infty$. The reported floor is rounded down. Replacing the rate grid by an interval-arithmetic bound on $K^*_u$, which is routine because $u$ is an explicit combination of exponentials, would make the certificate fully formal. \texttt{special\_branch.py} implements the convex program of \S\ref{sec:comp}. \texttt{exact\_witness.py} produces and verifies the exact realizations of Proposition~11; their rational data are stored in \texttt{exact\_N4.json}, \texttt{exact\_N5.json} and \texttt{exact\_N6.json}.

Stochastic complementation and censored chains are classical \citep{Meyer1989}. Earlier work on entropy production under coarse-graining and fast-state decimation includes \citet{PuglisiEtAl2010} and \citet{Jia2016}.

General separation-of-time-scales theory is well developed \citep{YinZhang2007}, and finite-state metastable reductions under rate-comparability assumptions are treated by Landim and Xu \citep{LandimXu2016}. In a different optimization problem, Dechant showed that prescribed probability evolution can be implemented with arbitrarily small entropy production by paying diverging activity \citep{Dechant2022}; this sharpens the contrast with the exact stationary path-law problem here, where Theorem 5 rules out rate escape at fixed state count. These works establish that fast-state reduction itself is not a new idea; the exact observational constraint, lower-semicontinuous closure, and attainment consequence are the formulation to audit here.

The equilibrium restriction on complete dwell densities is consistent with prior waiting-time approaches, including \citet{SkinnerDunkel2021}. Broader recent work on thermodynamic inference from coarse-grained trajectories and visible transitions is reviewed in \citet{Seifert2026Review}; structural inference of hidden paths is developed, for example, by \citet{MaierSeifertVanderMeer2025}. Lack of a reversible realization is not by itself a proof of a positive fixed-dimension gap above a positive all-dimension infimum. Skinner and Dunkel computed minimal entropy production compatible with waiting-time moments for hidden networks of fixed size, with the limit of large size taken numerically \citep{SkinnerDunkel2021}. Theorem~D is uniform in size by proof. The equilibrium form of complete dwell densities, (9.2), is classical \citep{Keilson1979}. The companion work \citep{AznagulovTwoPrices} uses the geometric-mean reversibilization \citep{DiaconisMiclo2009} in total variation, which gives linear dependence on the observed deviation and the sharp near-equilibrium order; Lemma~D.0 uses the arithmetic-mean companion in relative entropy, which tensorizes over sojourns. Inequalities between edge entropy production and symmetric divergences of the fluxes, such as the pseudo-entropy bound, appear in work on speed limits \citep{Dechant2022}. In the near-equilibrium regime, (D.2) reduces to a quadratic form in the edge forces. There the fluctuation--response relations of \citet{AslyamovPtaszynskiEsposito2025} are the natural tool for sharper certificates; we do not pursue this here.

The path-space argument uses the classical Skorokhod representation framework \citep{Skorokhod1956}. The Lipschitz infimum envelope in Theorem B is a standard variational construction; the substantive input here is the proved closure of the physical realization class.

The contributions we claim, concretely, are four: lower semicontinuity of the fixed-dimension realization cost under path-law perturbation, the exact state-priced observable representation, the architecture-selection consequences that follow from it, and the architecture-independent floor of Theorem~D with its consequences for $P_q$. None of the individual reduction tools is new. What we believe is new is proving them together in the exact-law, unrestricted-rate regime, and following the combination through to an attained, dual-representable optimum.

\section*{Code availability}
The exact-arithmetic verification script \texttt{verify\_thermodynamic\_realizability.py} and its generated JSON output are included with this preprint as ancillary files. The script reproduces the exact PH(4) construction and the stated diagnostic checks. It is reproducibility support only; the analytic proofs of Theorems 3--5 do not depend on it. The v2 scripts \texttt{floor\_certificate.py}, \texttt{special\_branch.py} and \texttt{exact\_witness.py}, with the exact data of Proposition~11, are included as well.

\bibliographystyle{unsrtnat}
\bibliography{thermo_references_audited}

\end{document}